\documentclass[a4paper,12pt,twoside]{article}

 \usepackage{ragged2e}
 \usepackage{graphicx}
 \usepackage{float}
 \usepackage{amsmath}
 \usepackage{mathtools}
\usepackage[numbers, square, comma, sort&compress]{natbib}
\usepackage{notoccite}
\usepackage{hyperref} 
\usepackage{authblk}
\usepackage{footmisc}
\usepackage{resizegather}    
\usepackage{latexsym}
\usepackage{amssymb,epsf}
\usepackage{titling}
\usepackage{geometry}
\usepackage{float}
 \setcitestyle{aysep={}}  

\begin{document}

  \begin{center}

	{\Large {\bf An interacting New Holographic Dark Energy in the framework of fractal cosmology} }
	
	\vspace*{12mm}
	{\large  Ehsan Sadri$^{a,}$\footnote{ehsan@sadri.id.ir}, Martiros Khurshudyan$^{b,c,d,e,}$\footnote{khurshudyan@yandex.com, khurshudyan@ustc.edu.cn, khurshudyan@tusur.ru} ,  Surajit Chattopadhyay$^{f,}$\footnote{schattopadhyay@kol.amity.edu}}
	\vspace*{0.5cm}
	
	{ {\it $^a$Department of Physics, Central Tehran Branch, Islamic Azad University, Tehran, Iran}} \\
	{ {\it $^b$International Laboratory for Theoretical Cosmology, Tomsk State University of Control Systems and Radioelectronics, 634050 Tomsk, Russia}} \\
	{ {\it $^c$Research Division, Tomsk State Pedagogical University, 634061 Tomsk, Russia}} \\
	{ {\it $^d$CAS Key Laboratory for Research in Galaxies and Cosmology, Department of Astronomy, University of Science and Technology of China, Hefei 230026, P. R. China}} \\
	{ {\it $^e$School of Astronomy and Space Science University of Science and Technology of China, Hefei 230026, P. R. China}}\\
{ {\it $^f$Department of Mathematics, Amity University, Major Arterial Road, Action Area II, New Town, Kolkata 700135, India}}
	
\vspace{15mm}

\end{center}

  \begin{abstract}
   In this paper, we study an interacting holographic dark energy model in the framework of fractal cosmology. The features of fractal cosmology could pass ultraviolet divergencies and also make a better understanding of the universe in different dimensions. We discuss a fractal FRW universe filled with the dark energy and cold dark matter interacting with each other. It is observed that the Hubble parameter embraces the recent observational range while the deceleration parameter demonstrates an accelerating universe and a behavior similar to $\Lambda$CDM. Plotting the equation of state shows that it lies in phantom region for interaction mode. We use $Om$-diagnostic tool and it shows a phantom behavior of dark energy which is a condition of avoiding the formation of black holes. Finally we execute the StateFinder diagnostic pair and all the trajectories for interacting and non-interacting state of the model meet the fixed point $\Lambda$CDM at the start of the evolution. A behavior similar to Chaplygin gas also can be observed in statefinder plane. We find that new holographic dark energy model (NHDE) in fractal cosmology expressed the consistent behavior with recent observational data and can be considered as a model to avoid the formation of black holes in comparison with the main model of NHDE in the simple FRW universe. It has also been observed that for the interaction term varying with matter density, the model generates asymptotic de-Sitter solution. However, if the interaction term varies with energy density, then the model shows Big-Rip singularity. Using our modified CAMB code, we observed that the interacting model suppresses the CMB spectrum at low multipoles $l<50$ and enhances the acoustic peaks. Based on the observational data sets used in this paper and using Metropolis-Hastings method of MCMC numerical calculation, it seems that the best value with $1\sigma$ and $2\sigma$
confidence interval are $\Omega_{m0}=0.278^{+0.008~+0.010}_{-0.007~-0.009}$,
$H_0=69.9^{+0.95~+1.57}_{-0.95~-1.57}$,
$r_c=0.08^{+0.02~+0.027}_{-0.002~-0.0027}$,
$\beta=0.496^{+0.005~+0.009}_{-0.005~-0.009}$,
$c=0.691^{+0.024~+0.039}_{-0.025~-0.037}$ and $b^2=0.035$
according to which we find that the proposed model in the presence of interaction is compatible with the recent observational data.
\end{abstract}


   \section{ INTRODUCTION}
  \justify
$~~~~$The universe is expanding with an accelerating rate since it entered in the dark energy dominated era. The accelerated expansion of the universe in charge of a notion in cosmology is an unsolved riddle. The dark energy as a negative-pressure fluid is the main reason for accelerated expansion of the universe\cite{{36},{37}}. There are many models proposed for dark energy to tackle the dark energy problems \cite{126},\cite{127},\cite{128},\cite{129},\cite{131},\cite{132},\cite{133},\cite{134},\cite{130} to mention a few. Following this, there are a large number of topics concerning the holographic dark energy models (HDE) discussed by cosmologists with various cosmological constraints. The Holographic dark energy model is originated from the holographic principle which is the most important foundation of quantum gravity and has a great potential to solve many issues of various physical fields discussed for a long time \cite{{40},{51},{52},{48},{59},{50},{54},{46},{56},{57},{47},{53},{32},{33},{61},{34},{58},{45},{55},{88}}. The Holographic dark energy model emphasizes on evaluation of number of degrees of freedom in a physical system including its bounding region rather than its volume\cite{38}. One of the main reasons of studying the holographic dark energy models is finding a way to prevent the formation of black holes (BHs) and better investigation of dark energy problems compared to other models \cite{118}, \cite{119}, \cite{120}, \cite{121}, \cite{122}, \cite{123}. In order to prevent the formation of black hole's mass, the minus sign of the equation of state of dark energy is not enough and the most suitable condition for satisfying this issue is a phantom-like dark energy \cite{{112},{66},{67},{63},{65},{64},{113},{114},{115}}. Despite the fact that the problem of the black hole's mass is not fully resolved, many authors have claimed that the phantom-like dark energy reduces the mass of black hole and it will be turned to be zero in this era before the Big Rip singularity\cite{{66},{67},{63},{65},{64},{142},{143}}. The mentioned Holographic dark energy models are particular forms of Nojiri and Odintsov model who checked the possibility of a universe with a phantom-like equation of state\cite{112}, \cite{113}, \cite{116}, \cite{117}. They also demonstrated that the viable dark energy for the inflationary early universe and late time acceleration of phantom like universe can be investigated by consideration of a covariant holographic dark energy and one can rewrite the modified gravity with scalar-tensor theory where the combination of the features of FRW universe is able to identify the infrared cutoff \cite{136}. Satisfying the conditions about preventing the formation of black hole's mass, a model has been proposed as pilgrim dark energy (PDE) \cite{72}. This model is a generalized holographic dark energy model which changes in its variable result in reaching different HDE models. Recently, this model (PDE) has been studied in the context of fractal cosmology and scrutinized its behavior towards avoiding the formation of BH\cite{73}. Getting benefit of the fractal framework's features, this work reached the fertilized condition to address the BH issue more than using the regular framework. There are also some other works in the field of the fractal cosmology which mentioned the phantom-like behavior in their context. A model of interacting dark energy with additional time dependent term has been purposed showing uncommon results regarding to other cosmological models\cite{74}. The other paper which worked in this framework\cite{75} studied the generalized HDE and dynamical characteristic of its potential. Furthermore, \cite{86} studied the new agegraphic and the ghost dark energy models in the fractal cosmology and investigated the behavior of the equation of state. \cite{87} studied the thermodynamic features of the apparent horizon in the framework of fractal universe. \cite{124} studied the nonlinear interacting dark energy model in this framework and \cite{125} investigated the fractal analysis toward the distribution of galaxies. \newline
$~~~~$It is worthwhile to explain the fractal framework of the universe. According to the profound relation between gravity and thermodynamic, using the Jacobson's derivation of Einstein gravity, one has a new way to comprehend the thermodynamics behavior of gravity\cite{89}. Using this, many authors have had efforts to show that the gravitational field equations can be written in the form of first law of thermodynamics \cite{90}, \cite{91}, \cite{92}, \cite{93}. Regarding this matter, it is also shown that the Friedmann equation in the FRW cosmology can reach the first law of thermodynamics, even by the use of the fractal cosmology as another theoretical approach \cite{87}. In physical cosmology, the fractal cosmology is introduced as a community of theories stating about the fractality over a comprehensive range of scales of the universe and the matter distribution inside it. In large or even small scales, the fractal dimension or matter distribution of the universe is very important. The fractal cosmology\cite{{1},{2}} - discussed by \cite{95} for the first time - is a power-counting renormalizable field theory existing in a fractal space-time and without ultraviolet (UV) divergence. The renormalizability in theory leads to stable ultraviolet\cite{98}, \cite{101}. In addition, in this framework in the vicinity of two topological dimensions, the renormalizability of perturbative quantum gravity theories lead to more attention to $D=2+\epsilon$ models being able to improve our understanding of four dimensional cases ($D=4$). \cite{96}, \cite{97}, \cite{98}, \cite{99}, \cite{100}. For further information, using the fractal framework can be for this reason that our universe witnesses fractals on many levels. One can see the fractal characteristic of the quantum gravity in $D$ dimensional, for $D=3$ and $D<3$ resulting the regular galaxy distribution and inhomogeneous galaxy distribution, respectively \cite{102}, \cite{103}, \cite{104}.\\
$~~~$On the other hand, the deep connection between gravitational terms being described in the bulk and first law of thermodynamic can lead to various ideas of holography. Following this claim, in recent years brane theory tiding up in a higher dimensional space-time has drawn many attention\cite{108}, \cite{109}, \cite{110}, \cite{111}. In these theories, cosmic evolution can be explained by a Friedmann equation interacting with the bulk's effect on the brane. The most popular model in the framework of the braneworld has been proposed as DGP which stands for DvaliGabadadze-Porrati\cite{105}. The DGP model changes the four dimensional universe to five dimensional minkovskian bulk. The self-accelerating feature of DGP model has the ability of conveying the late time acceleration of the universe free from any dark energy relation \cite{106}, \cite{107}. This feature of DGP model cannot satisfy the phantom line crossing and for this issue adding an energy term on the brane is required \cite{8}. Accordingly, this added energy term would lead to emergence of a novel way for explanation of the late time acceleration and better compatibility with observational data\cite{8}.\\
$~~~$Given the aforementioned explanations, in this paper, we would like to study a new model of holographic dark energy (NHDE) based on DGP braneworld in the framework of fractal cosmology\cite{8}. In spite of the mentioned reasons for this choice of HDE, it may note that this model is very particular example of general HDE introduced by\cite{113}, where a phantom cosmology towards unification of early and late time universe was proposed. We also study the initial matter power spectrum and CMB angular power spectrum are generated by our modified Einstein-Boltzmann CAMB code \cite{138}, \cite{139} with the coupling between dark energy and dark matter which is different from the $\Lambda$CDM model in low-$l$ with interacting model. We will study the relevant analysis in this paper for other generalizations of HDE and their comparison as an important issue to find out the behavior of them as future work.

The structure of this paper is as follows. In the next section we review the main equations of the fractal cosmology. In section 3 we study New Holographic Dark Energy Models in the framework of the fractal cosmology. In section 4,  we use diagnostic tools $s-r$ and $Om$ planes to characterize properties of dark energy. In section 5, we study the behavior of present model to understand the type of future singularity. In section 6, we study the behavior of the present model in CMB angular power spectrum and matter power spectrum plots. Finally, by use of the combination of recent observational data sets (SN Ia + BAO + CMB + OHD) we fit the relevant free parameters. For this issue, we used Markov Chain Monte Carlo (MCMC) method. The last section is allocated to some concluding remarks.

    \section{ FRACTAL COSMOLOGY}
    The total action of Einstein gravity in a fractal space-time $(S=S_G+S_m)$ is given by\cite{{1},{2}}
    \begin{equation}
   S_G=\frac{1}{16\pi G}\int d\varrho\left(x\right)\sqrt{-g}\left(R-2\Lambda-\omega\partial_{\mu}\nu\partial^{\mu}\nu\right)\
   \end{equation}
   and the matter action is
   \begin{equation}
  S_m=\frac{1}{16\pi G}\int d\varrho\left(x\right)\sqrt{-g}\mathcal{L}_m\
  \end{equation}
    with respect to the Friedmann-Robertson-Walker (FRW) metric $g^{\mu\nu}$ one can obtain the Friedmann equation in a flat fractal universe as following\cite{2}
    \begin{equation}\label{frwfractal}
  H^2+H\frac{\dot{\nu}}{\nu}-\frac{\omega}{6}\dot{\nu}^2=\frac{1}{3}\rho+\frac{\Lambda}{3}
   \end{equation}
   where $\rho=\rho_D+\rho_m$ and $H=\frac{\dot{a}}{a}$
      The continuity equation in a fractal universe takes the form\cite{2}
   \begin{equation}\label{continuity}
\dot{\rho}+\left(3H+\frac{\dot{\nu}}{\nu}\right)\left(\rho+P\right)=0
   \end{equation}
   In the framework of fractal cosmology, fractals can be time-like ($\nu(t)$) or space-like ($\nu\left(x\right)$)\cite{2}. In this paper we choose a time-like fractal in order to work on functions of scale factor or redshift. Hence, all parameters related to time change to be appropriate in our calculations. Consideration of a timelike fractal profile\cite{2} as $\nu=t^{-\beta}$ would lead to a divergent production of energy as  $t \rightarrow 0$. For this, with assumption of a well behaved approximation of the universe in the whole expansion at large scales ($a(t)\approx t$) one can pass this divergency and we have
    \begin{equation}\label{timelike}
  \nu=a^{-\beta}
   \end{equation}
   where $a$ is scale factor and $\beta=D(1-\alpha)$. Choosing timelike fractal profile, the UV regime explains short scales resulting that inhomogeneties\footnote[4]{In inhomogeneous cosmology the solutions of the Einstein field equations makes the large scale structure of the universe while in the theory of cosmological perturbations being the study of the Universe in a perturbative way, structure formation will be considered\cite{135}} play some role. If these scales are small, the modified Friedmann equations define a background for perturbations rather than a self-consistent dynamics Using \ref{timelike} changes Eq. \ref{frwfractal} to
    \begin{equation}\label{fractal_H}
   H^2\left(1-\beta-\frac{\beta^2\omega a^{-2\beta}}{6}\right)=\frac{1}{3}\left(\rho_D+\rho_m\right)
   \end{equation}
where $D=4$ which stands for four-dimensional space and $0<\alpha\leqslant1$\cite{{3},{2}}. Now conservation Eq. \ref{continuity} trasforms to
 \begin{equation}\label{Q}
  \dot{\rho}_m+\left(3-\beta\right)H \rho_{m}=Q
   \end{equation}
    \begin{equation}\label{-Q}
  \dot{\rho}_D+\left(1+\omega_D\right)\left(3-\beta\right)H\rho_D=-Q
   \end{equation}
 where $\rho_D$ and $\rho_m$ are densities of dark matter and dark energy respectively, $\omega_D$ is the equation of state parameter for dark energy. The behavior of interaction with different Q-terms have been studied in\cite{{4},{5},{7},{6}}. In this work we take a simple interaction term as ($Q=3b^2H\rho_m$) which explains an interaction between dark energy and cold dark matter. Using this interaction term, the conservation Eqs. \ref{Q} and \ref{-Q}  in this case take the forms
         \begin{equation}\label{Qopen}
  \dot{\rho}_m=\left(3\left(b^2-1\right)+\beta\right)H\rho_{m_0}a^{3\left(b^2-1\right)+\beta}
   \end{equation}
    \begin{equation}\label{-Qopen}
  \dot{\rho}_D=-\left(3-\beta\right)\left(1+\omega_D\right)\rho_DH-3b^2H\rho_{m_0}a^{3\left(b^2-1\right)+\beta}
   \end{equation}
    where $b^2$ is the coupling constant and $\rho_{m_0}$ is the present value of dark matter density.\\

   \section{ NHDE IN FRACTAL COSMOLOGY }

   The energy density of the new holographic dark energy (NHDE) is given by the following relation\cite{8}
\begin{equation}\label{NHDE}
   \rho_D=\frac{3c^2}{ L^2}\left(1-\frac{\epsilon L}{3 r_c}\right)
   \end{equation}
   in which $r_c=(2H\sqrt{\Omega_{r_c}})^{-1}$ is the crossover length scale, $\epsilon = \pm1$ related to the two answers of solution\cite{106} and $L=H^{-1}$ is Hubble horizon as the system's IR cutoff. For $\epsilon=+1$, the universe lies within an accelerating phase in the late time with no further dark energy element. As $L\ll 3r_c$, Eq.\ref{NHDE} reduces to the main holographic dark energy density. Taking time derivative of relation \ref{NHDE} leads us to
     \begin{equation}\label{rhod}
   \dot{\rho}_D=3c^2\left(H^3\left(1-\frac{\epsilon}{3r_cH}\right)+\frac{\epsilon H^2}{3r_c}\right)\frac{\dot{H}}{H^2}
   \end{equation}
    \begin{figure}[H]
\centering
\includegraphics[scale=0.44]{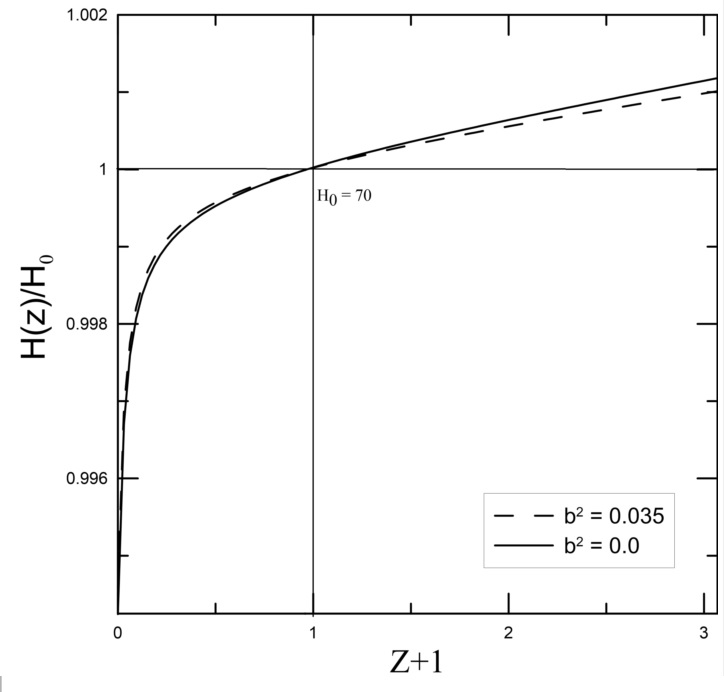}
\caption{The evolution of Hubble parameter ($H/{H_0}$).}
\label{fig1}
\end{figure}
     Taking time derivative of Eq. \ref{fractal_H} and using Eq.. \ref{Qopen}, \ref{NHDE} and \ref{rhod} yields
    \begin{equation}\label{HdH2}
 \frac{\dot{H}}{H^2}=\frac{\frac{1}{H^2}\left(3b^2-3+\beta\right)\rho_{m_0}a^{3\left(b^2-1\right)+\beta}-\beta^3\omega a^{-2\beta}}{6\left(1-\beta-\frac{\beta^2\omega a^{-2\beta}}{6}-c^2\left(1-\frac{\epsilon}{2r_cH}\right)\right)}
   \end{equation}
   where $\rho_{m_0}$ is the integration constant in Eqs. \ref{Qopen} and\ref{-Qopen} and $\Omega_{m_0}=\frac{\rho_{m_0}}{3H_0}$ is the matter density parameter of present time. Solving Eq. \ref{HdH2} with some considerations $d/dx=-(1+z)d/dz$ and $H'=\dot{H}/H$, one can check the evolution of Hubble parameter versus redshift  as seen in Fig. \ref{fig1}.  In the flat universe ($\Omega_k=0$), using $\Omega_m+\Omega_D=1-\frac{\epsilon}{2r_cH(z)}$, $\Omega_D=c^2(1-\frac{\epsilon}{6r_cH(z)})$ and Eq.\ref{HdH2}, we can plot Fig. \ref{fig1}. This Fig. demonstrates a good compatibility with observational data \cite{140}, \cite{141}.
In order to understand the behavior of the current model, using Eq. \ref{-Qopen} we can extract the EoS parameter
\begin{equation}\label{eos1}
 \omega_D=-1-\frac{3b^2H\rho_{m_0}a^{3\left(b^2-1\right)+\beta}-\dot{\rho}_D}{\left(3-\beta\right)H\rho_D}
   \end{equation}
   The combination of Eqs. \ref{NHDE}, \ref{rhod}  and \ref{eos1} yields
   \begin{equation}\label{eos2}
  \omega_D=-1-\left(\frac{3b^2H\rho_{m_0}a^{3\left(b^2-1\right)+\beta}}{H^2}\right)\left( \left(3\left(3-\beta\right)c^2\left(1-\frac{\epsilon}{3r_cH}\right)\right)^{-1} -\left(2+\frac{\epsilon}{3r_cH\left(1-\frac{\epsilon}{3r_cH}\right)}\right)\right)\left(3-\beta\right)^{-1}\frac{\dot{H}}{H^2}
   \end{equation}
Using Eq. \ref{HdH2} into \ref{eos2} we plot the right plan of Fig. \ref{fig2}. The deceleration parameter can be explained as
  \begin{equation}
   q=-\frac{a\ddot{a}}{\dot{a}^2}=-1-\frac{\dot{H}}{H^2}
   \end{equation}
   and using Eq. \ref{HdH2} we have
     \begin{equation}\label{dec}
      q=-1-\left(\frac{1}{H^2}\left(3b^2-3+\beta\right)\rho_{m_0}a^{3\left(b^2-1\right)+\beta}-\beta^3\omega a^{-2\beta}\right)\left(6\left(1-\beta-\frac{\beta^2\omega a^{-2\beta}}{6}-c^2\left(1-\frac{\epsilon}{2r_cH}\right)\right)\right)^{-1}
       \end{equation}
   The evolutions of Hubble parameter, Deceleration parameter and equation of state against redshift are plotted in Fig. \ref{fig1}and \ref{fig2} respectively. We find $b^2$ as a positive small value using observational data which it is mentioned in \cite{{11},{10}} as well.
\begin{figure}[H]
   \centering
\begin{tabular}{ccc}
\hspace*{-0.1in}
\includegraphics[width=0.48\textwidth]{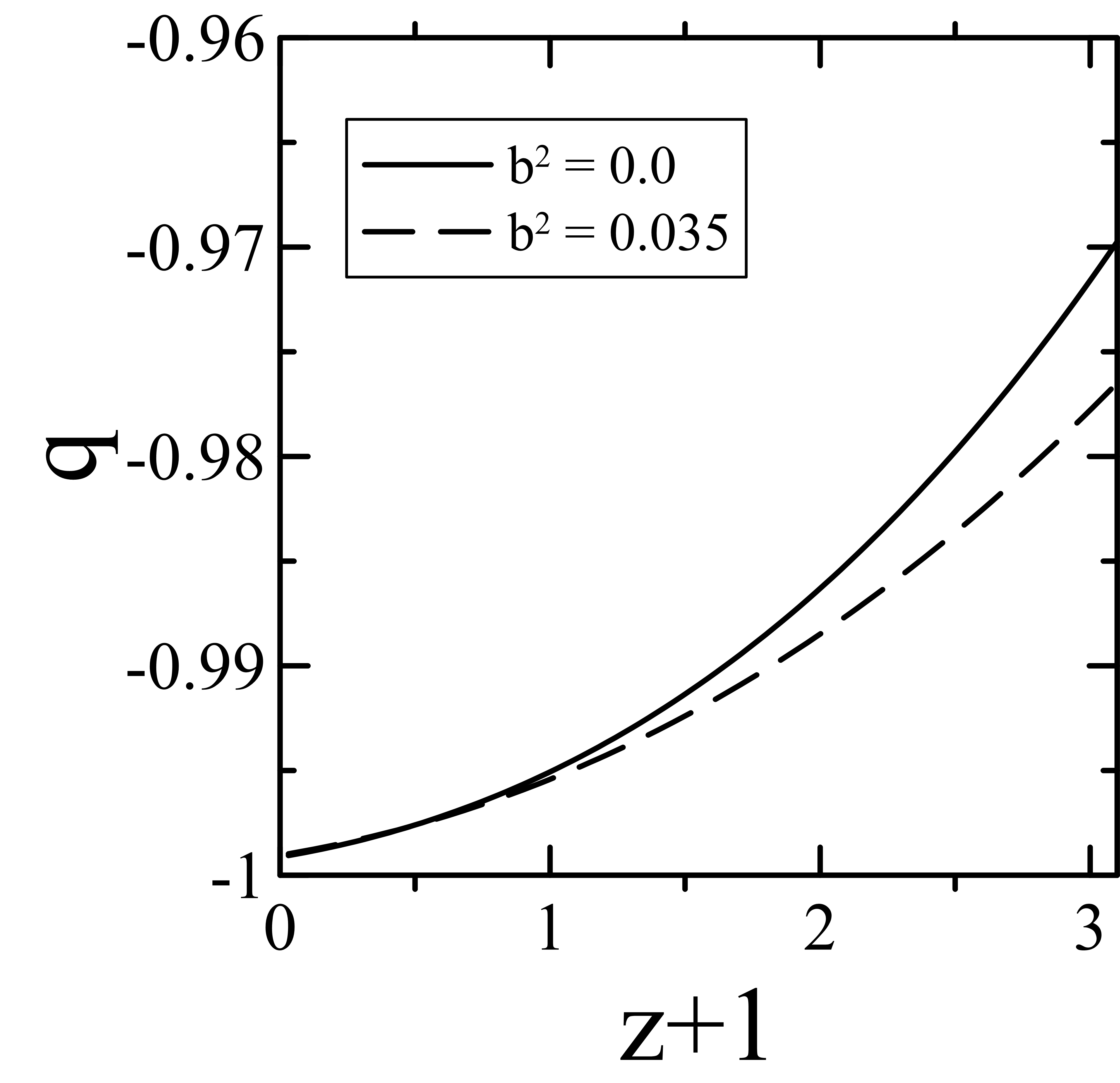}\hspace{2mm}
\includegraphics[width=0.46\textwidth]{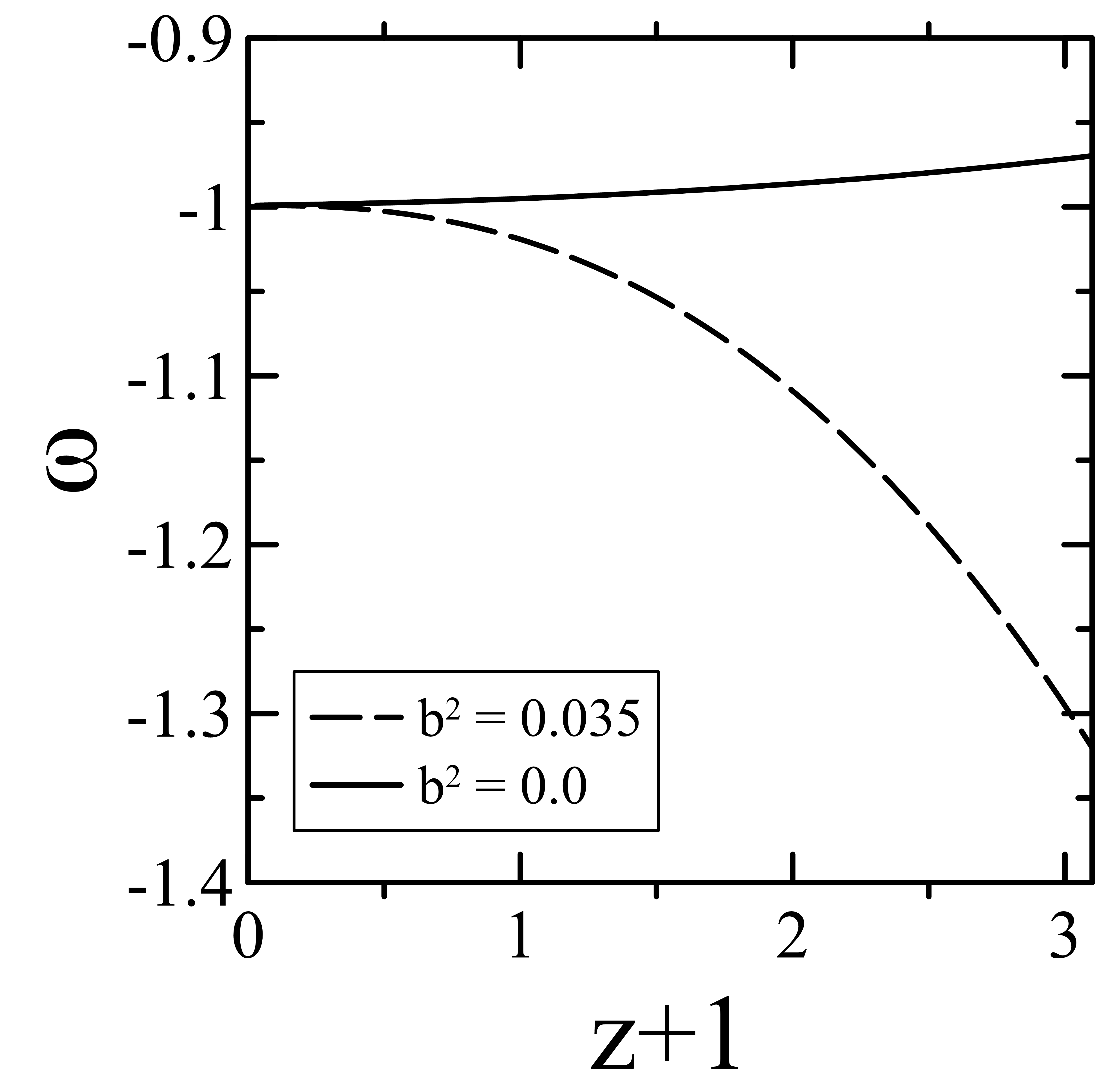}\end{tabular}
\caption{The evolution of the equation of state and the deceleration parameter in terms of redshift} \label{fig2}
\end{figure}
In Fig. \ref{fig2}, the deceleration parameter is very close to $-1$ which denotes a behavior similar to cold dark matter and shows the universe with accelerated rate of the whole expansion.The equation of state also just lies in phanton realm ($\omega_D<-1$) since the model wintnesses the interaction between dark components $b^2>0$.\\

   \section{DIAGNOSTIC RECOGNITION}
In this section we propose two tracing tools to check the characteristic of the present Dark energy model.

\subsection{$Om$-DIAGNOSTIC}
\begin{figure}[H]
\centering
\includegraphics[scale=0.44]{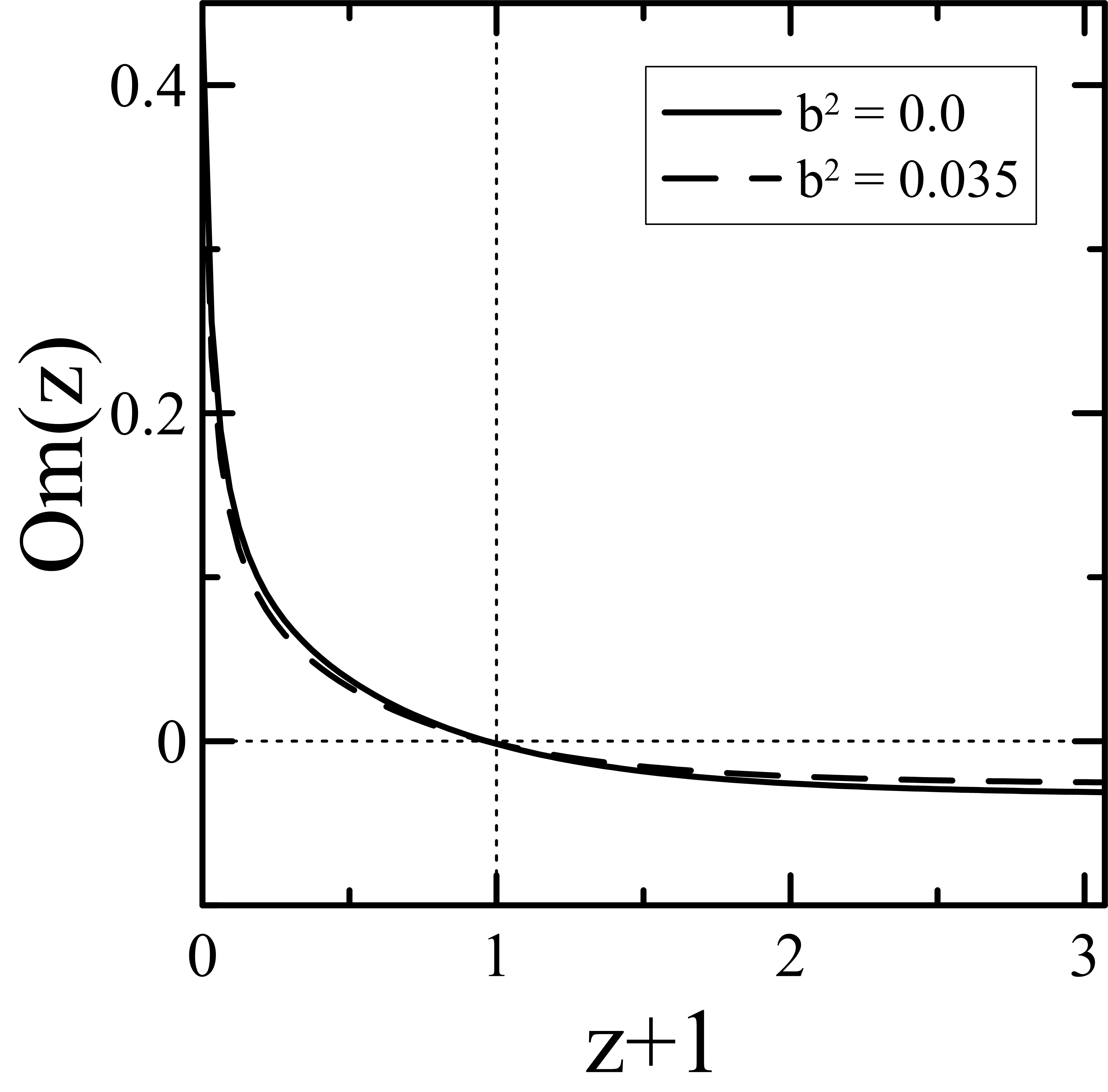}
\caption{The evolution of $Om$-diagnostic tool versus $z$}
\label{fig4}
\end{figure}

 To check different periods of the universe, the $Om$-diagnostic tool has been proposed\cite{15}. The behavior of DE model can be discriminated by the use of this dignostic tool and also according to resulted trajectories in the final plot. This plot is divided into two parts. Phantom-like ($\omega_D< −1$) for the positive trajectories of $Om(x)$ and quintessence ($\omega_D>−1$) for the negative value of trajectories. The Om-diagnostic tool may be defined as
    \begin{equation}
Om\left(x\right)=\frac{h\left(x\right)^2-1}{x^3-1}
   \end{equation}
where $h(x)=H(x)/H_0$ and $x=ln(z+1)^{-1}$. The evolvement of $Om$-diagnostic tool versus redshift is depicted in Fig.\ref{fig4}. It is obvious that in the late time the trajectories for interacting and non-interacting mode present positive values conveying phantom like behavior and emphasizing on the EoS parameter result as seen in Fig.\ref{fig2}. However, in $z+1>1$, the trajectories demonstrate negative values which implies an universe with quintessence behavior.

\subsection{STATEFINDER DIAGNOSIS PAIR}
\begin{figure}[H]
\centering
\includegraphics[scale=0.44]{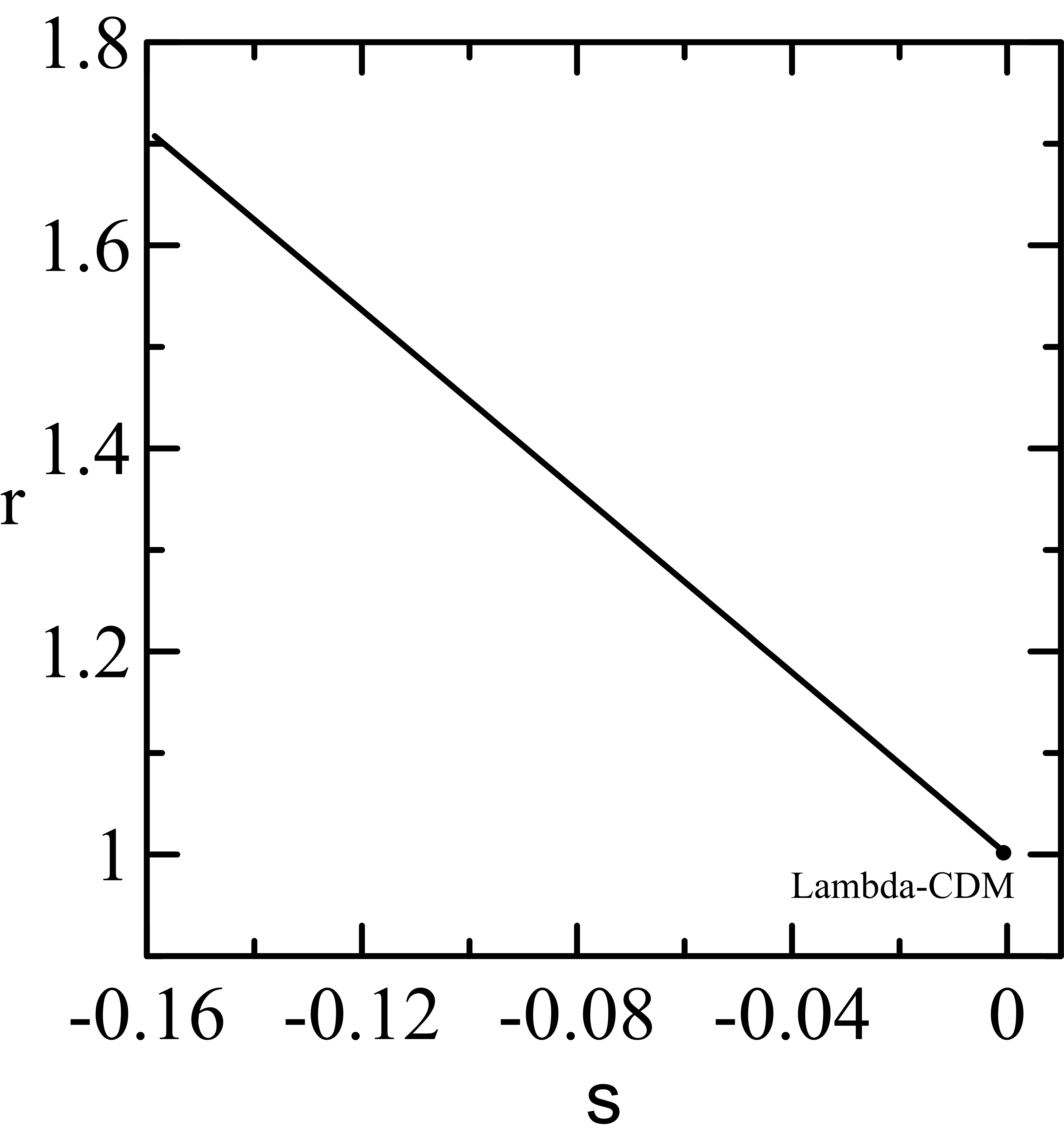}
\caption{The evolution of the StateFinder pair}
\label{fig5}
\end{figure}

In spite of the fact that the evolution of cosmic expansion is defined by the Hubble parameter (H) and rate of acceleration and deceleration of this expansion is explained by Eq. \ref{eos2} and \ref{dec}, we cannot clearly differentiate various dark energy models using these two parameters when $H > 0$ or $q < 0$. Hence, to do proper calculations about distinguishing the various dark energy models and due to the development of observational data during the recent two decades a new geometrical diagnostic pair has been proposed \cite{{16},{17}} which let us to specify the features of dark energy. This new pair is called StateFinder pair$(r,s)$
     \begin{equation}\label{rs}
r=\frac{\dddot{a}}{aH^3}=1+\frac{\ddot{H}}{H^3}+3\frac{\dot{H}}{H^2}~~~~~~~~~~~~~~~~~s=\frac{r-1}{3\left(q-\frac{1}{2}\right)}
   \end{equation}
For investigating statefinder \ref{rs} for NHDE in the framework of fractal cosmology, we must have $\frac{\ddot{H}}{H^3}$. Consequently we can calculate ($s$). Taking the time derivative of both sides of Eq. \ref{HdH2} we have
    \begin{equation}\label{hdd}
    \begin{split}
\frac{\ddot{H}}{H^3}=\left(-\frac{2}{3}\frac{\dot{H}}{H^2}\frac{\left(3B^2-3+\beta\right)\rho_{m_0}a^{3\left(b^2-1\right)+\beta}}{H^2}+\frac{\alpha \left(3B^2-3+\beta\right)\rho_{m_0}a^{3\left(b^2-1\right)+\beta}}{H^2}+\frac{2\beta^4 \omega a^{-2\beta}}{3}\right)~~~~~~~~~~\\
\times\left(2\left(1-\beta-\frac{\beta^2\omega a^{-2\beta}}{6}\right)-2c^2\left(1-\frac{\epsilon\sqrt{\Omega_{r_c}}}{3}\right)\right)^{-1}-\left(\frac{\beta^3\omega a^{-2\beta}}{3}-\frac{2c^2\sqrt{\Omega_{r_c}}\dot{H}}{3}\right)\left(\frac{1}{3H^2}-\left(\frac{\beta^3\omega a^{-2\beta}}{3}\right)\right)\\
\times\left(2\left(1-\beta-\frac{\beta^2\omega a^{-2\beta}}{6}\right)-2c^2\left(1-\frac{\epsilon\sqrt{\Omega_{r_c}}}{3}\right)\right)^{-2}+2\left(\frac{\dot{H}}{H^2}\right)^2~~~~~~~~~~~~~~~~~~~~~~~~~~~~~~~~~~~~~~~~~~~~~~~~~~~
\end{split}
   \end{equation}
   Using Eqs. \ref{HdH2}, \ref{dec} and \ref{hdd} into \ref{rs} we plot Fig. \ref{fig5}. In Fig.\ref{fig5}, we can see the evolution of trajectories where the horizontal axis and vertical axis are defined by parameter $s$ and $r$ respectively. As the universe expands, it is interesting to note that the parameter $r$ increases while the parameter $s$ decreases. The evolution of trajectories advance from the positive part of $s$ parameter to negative. In addition, the plot is close to the $\Lambda$CDM fixed point area $(r, s) = (1,0)$ for both interacting and non-interacting form $(r=1.0035, s=-0.0007)$. The StateFinder trajectory demonstrates the Chaplygin gas behavior (where $s<0$, $r>1$). Moreover, it shows that $s<0$ corresponding to a phantom-like Dark energy behavior. This is an affirmation on the equation of state results.

\section{SINGULARITY}

Based on \cite{83} we consider the Hubble
parameter $H$ as
\begin{equation}\label{singularity}
H=\left(\frac{h}{t_0-t}\right)^\beta+H_0
\end{equation}
 to understand the type of future
singularity it will lead to for the present case.

At this juncture it may be noted that for the sake of simplicity
without any loss of generality, we have chosen $\beta$ in the
above choice of $H$ and this coincides with $\beta=D(1-\alpha)$ in
Eq. \ref{timelike} and this may be considered as a suitable choice as four of
the types of signularity discussed below can have feasible ranges
of $\beta$ in consistency with Eq. \ref{timelike}. The finite-time future
singularities can be classified as follows \cite{83}:
\begin{itemize}
    \item Type I (Big Rip) \cite{84}: for
$t\rightarrow t_0$ , $a(t)\rightarrow\infty$,
$\rho_{eff}\rightarrow \infty$, $|p_{eff}|\rightarrow \infty$.
This corresponds to $\beta> 1$ and $\beta=1$. This also includes
the cases in which $\rho_{eff}$ and $p_{eff}$ are finite at $t_0$
is also included.
    \item Type II (sudden): for
$t\rightarrow t_0$ , $a(t)\rightarrow a_0$, $\rho_{eff}\rightarrow
\rho_0$, $|p_{eff}|\rightarrow \rho_0$. It corresponds to $- 1 <
\beta < 1$.
    \item Type III: for
$t\rightarrow t_0$ , $a(t)\rightarrow a_0$, $\rho_{eff}\rightarrow
\infty$, $|p_{eff}|\rightarrow \infty$. This corresponds to $0 <
\beta < 1$.
    \item Type IV: for
$t\rightarrow t_0$ , $a(t)\rightarrow a_0$, $\rho_{eff}\rightarrow
0$, $|p_{eff}|\rightarrow 0$. and higher derivatives of $H$
diverge. It also includes the case of $\rho$ and or $p$ tending to
finite values. It corresponds to $\beta < - 1$ and $\beta$ is not
an integer.
\end{itemize}
For the choice of Hubble parameter as in Eq. (\ref{singularity})
the scale factor comes out to be
\begin{equation}\label{scalefactor}
a\left(t\right)=C_1 e^{H_0
t+\frac{h\left(t_0-t\right)^{1-\beta }}{-1+\beta }}
\end{equation}
Considering the interaction term as $Q=3b^2H\rho_D$ i.e.
proportional to the dark energy density, using Eq. \ref{rhod} the EoS
parameter is found to be

\begin{equation}\label{eosd1}
w_D=\left(-3-3 b^2 \left(H_0+h \left(t_0-t\right)^{-\beta }\right)+\beta
+\frac{6 C^2 h \left(h+H_0 \left(t_0-t\right)^{\beta }\right) \beta }{\left(t-t_0\right)
\left(-2 C_2 \left(t_0-t\right)^{2 \beta }+3 C^2 h \left(h+2 H_0
\left(t_0-t\right)^{\beta }\right)\right)}\right)\left(3-\beta\right)^{-1}
\end{equation}
In this scenario, if $t\rightarrow t_0$ and $\beta>1$, then
$a(t)\rightarrow \infty$ and also Eq. (\ref{eosd1}) makes it
apparent that $w_D\rightarrow \infty$. In the current scenario we
also have using Eq. \ref{rhod} that
\begin{equation}
\rho_D=-\frac{3}{2} c^2 h \left(t_0-t\right)^{-2 \beta } \left(h+2 H_0
\left(t_0-t\right)^{\beta }\right)+C_2
\end{equation}
and
\begin{equation}
 \begin{split}
p_D=\left(-3-3 b^2 \left(H_0+h \left(t_0-t\right)^{-\beta
}\right)+\beta +\frac{6 c^2 h \left(h+H_0 \left(t_0-t\right)^{\beta }\right)
\beta }{\left(t-t_0\right) \left(-2 C_2 \left(t_0-t\right)^{2 \beta }+3 c^2 h \left(h+2
H_0 \left(t_0-t\right)^{\beta }\right)\right)}\right)\\
\times\left(\left(3-\beta \right)\left(C_2-\frac{3}{2} c^2 h \left(t_0-t\right)^{-2 \beta }\left(h+2 H_0
\left(t_0-t\right)^{\beta }\right)\right)\right)^{-1}~~~~~~~~~~~~~~~~~~~~~~~~~~~~~~~~~~~~~~~~~~~~~~~~
\end{split}
\end{equation}
Clearly as $t$ tends to $t_0$, the $\rho_D$ tends to infinity for
$\beta=1$ as well as $\beta>1$. As an obvious consequence
$\rho_{eff}$ would tend to $\infty$. Also, under the same
constraints, $p_{eff}\rightarrow 0$. In view of the discussion
presented above, it may be interpreted that for interaction term
$Q$ chosen proportional to the dark energy density, the model is
characterized by Type I i.e. Big Rip singularity. Hence, according to the consideration of general choice for $H$ we observe that the present work conveys the Big Rip singularity as the Type 1.
Let us now consider the interaction term proportional to $\rho_m$,
i.e. $Q=3b^2H\rho_m$ as already considered in Section 2 and all
the subsequent cosmological parameters are constructed based on
this choice. Through simple computation it can be shown that for
the present choice of Hubble parameter, $\dot{H}/H^2$ and $1/H$
tend to $0$ as $t\rightarrow t_0$ and $\beta>1$ or $\beta=1$. If
we use this in Eq. \ref{eos2} we observe that $w_D\rightarrow -1$ as
$t\rightarrow t_0$ and hence we get the asymptotic de Sitter
universe. This solution corresponds to the so called
w-singularity. This type of solution was obtained in Astashenok et
al. \cite{85}, where they constructed phantom energy models
with the equation of state parameter $w$ such that the equation of
state parameter tends to constant value with time
i.e.``cosmological constant" with asymptotically de Sitter
evolution. Hence, for $Q=3b^2H\rho_m$, we have asymptotically de
Sitter evolution in the scenario where NHDE is considered in the
framework of fractal cosmology. A detailed account on this issue is discussed in \cite{144}.
\section{Power Spectrum}
\begin{figure}[H]
\centering
\includegraphics[scale=0.5]{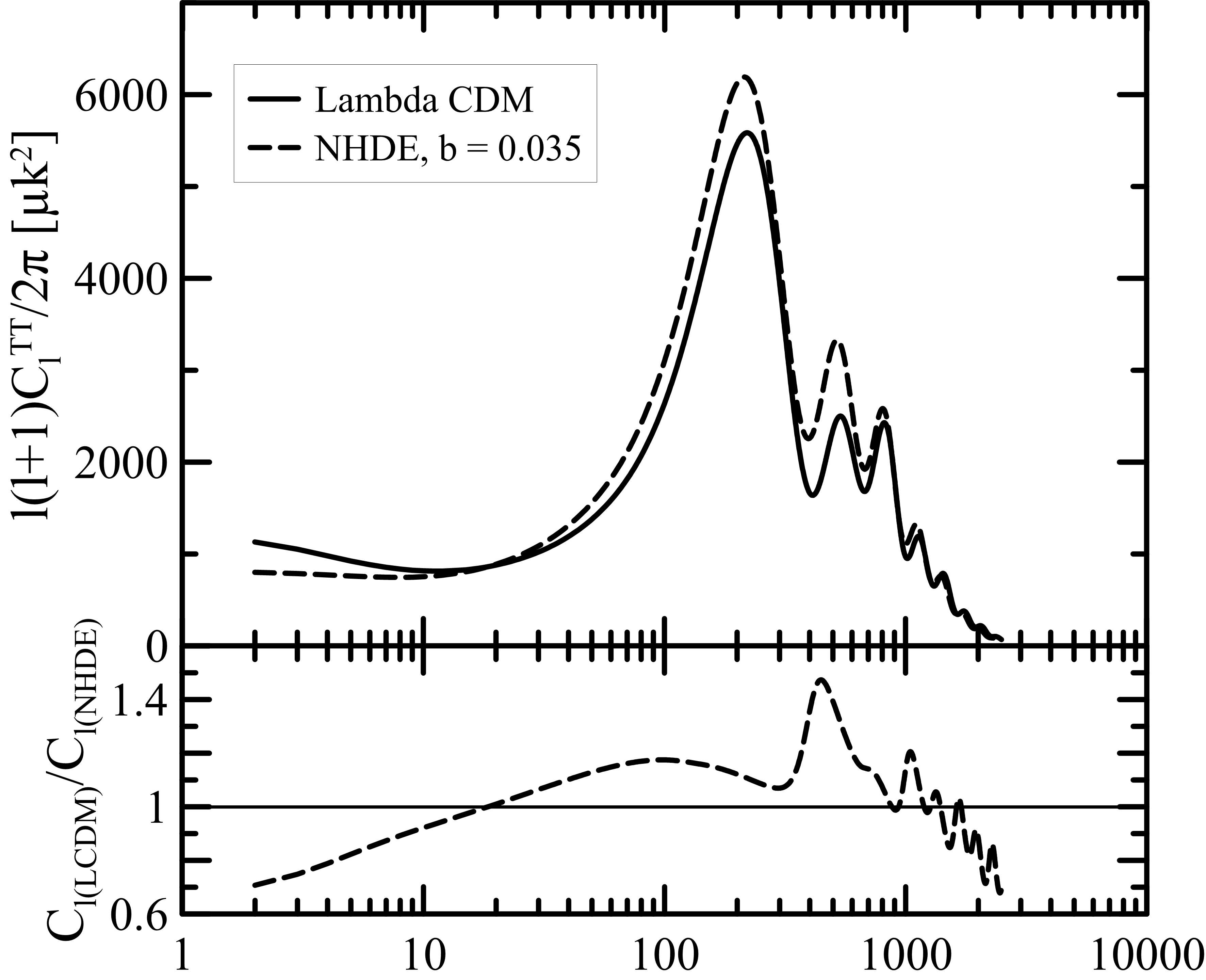}
\caption{The CMB temperature spectra $c_l^{TT}$ with interacting and non-interacting model of NHDE in the fractal universe. Here we set the decoupling constant known as the interaction component, $b=0$ for non-interacting and $b=0.035$ for interacting model. The plot of non-interacting model coincide the $\Lambda$CDM model and behave similar to cold dark matter. As we can see, the interacting model leads to suppression of the CMB temperature at $l<50$.}
\label{figcl}
\end{figure}

In this section by the use of modified version of the Boltzmann code CAMB\footnote{https://camb.info/} \cite{138}, \cite{139} we compare the present model with $\Lambda$CDM model. In the Fig. \ref{figcl} the temperature power spectrum of the best fitted parameters of NHDE model is depicted. In this case, the interaction between dark sectors results in the enhancement of the acoustic peaks. The interacting NHDE behaves similar to $\Lambda$CDM, but the main difference between the NHDE model and $\Lambda$CDM model lies before $l<50$. In this area, we find that the amplitude of NHDE is lower than $\Lambda$CDM which means interacting terms suppress CMB spectrum at low multipoles due to the late ISW effets. This behavior is consistent with \cite{137} showing a phantom-like behavior of the model. The phantom-like behavior leads to smaller $c_l^{TT}$ at low-$l$ area.

\section{Data Analysis Methods}
Using the latest observational data including SN Ia, BAO, CMB and OHD, we constrain the free parameters of current model to obtain the best fit values. For this purpose, we use the minimized chi-square test for $1\sigma$ and $2\sigma$ confidence area.\\
For the Supernova type Ia (SNIa), we use the compressed Joint Light Analysis (cJLA) data set of 30 binned check points with the range of redshift $z=[0.01,1.3]$ \cite{77}. The $\chi^2$ function for SNIa having an accurate approximation of the full JLA likelihood is
\begin{equation}\label{SN}
\chi^2_{SNIa}=r^{t} C^{-1}_b r,
\end{equation}
in which
\begin{equation}\label{rSNIa}
r=\mu_b-M-5log_{10}d_L,
\end{equation}
and $C_b$ covariance matrix of $\mu_b$ \cite{77} is

\setcounter{MaxMatrixCols}{31}\label{rSNIa}

\begin{gather*}
  \setlength{\arraycolsep}{.25\arraycolsep}
  \text{\large$\displaystyle
10^{-6}\begin{pmatrix}  21282& -10840& 1918& 451& 946& 614& 785& 686& 581& 233& 881& 133& 475& 295& 277& 282& 412& 293& 337& 278& 219& 297& 156& 235& 133& 179& -25& -106& 0& 137& 168\\
&28155& -2217& 1702& 74& 322& 380& 273& 424& 487& 266& 303& 406& 468& 447& 398& 464& 403& 455& 468& 417& 444& 351& 399& 83& 167& -86& 15& -2& 76& 243\\
&&6162& -1593& 1463& 419& 715& 580& 664& 465& 613& 268& 570& 376& 405& 352& 456& 340& 412& 355& 317& 341& 242& 289& 119& 152& -69& -33& -44& 37& 209\\
&&&5235& -722& 776& 588& 591& 583& 403& 651& 212& 555& 353& 355& 323& 442& 319& 372& 337& 288& 343& 210& 272& 92& 167& -48& -29& -21& 50& 229\\
    &&&&7303& -508& 1026& 514& 596& 315& 621& 247& 493& 320& 375& 290& 383& 286& 350& 300& 269& 313& 198& 251& 99& 126& 18& 46& 13& 10& 203\\
             &&&&                      &3150& -249& 800& 431& 358& 414& 173& 514& 231& 248& 221& 293& 187& 245& 198& 175& 231& 126& 210& 103& 170& 51& 66& -8& -51& 308\\
                      &&&&&                                    &3729& -88& 730& 321& 592& 188& 546& 316& 342& 290& 389& 267& 341& 285& 252& 301& 189& 242& 122& 159& 35& 72& 30& 28& 255\\
                                  &&&&&&                                 &3222& -143& 568& 421& 203& 491& 257& 280& 240& 301& 221& 275& 227& 210& 249& 148& 220& 123& 160& 43& 69& 27& 7& 253\\
                                                &&&&&&&                            &3225& -508& 774& 156& 502& 273& 323& 276& 370& 260& 316& 273& 231& 273& 171& 226& 111& 154& 0& 29& 19& 23& 206\\
                                                                &&&&&&& &               &5646& -1735& 691& 295& 362& 316& 305& 370& 280& 346& 313& 276& 310& 217& 274 &131&175& 38& 118& 78& 48& 303\\
                                                                                &&&&&&&&&       &8630& -1642& 944& 152& 253& 184& 274& 202& 254& 233& 196& 237& 156& 207& 27& 115& -32& 7& -15& 0& 176\\
                                                                                                     &&&&&&&&&&  &3855& -754& 502& 225& 278& 294& 74& 285& 253& 239& 255& 173& 229& 181& 177& 93& 124& 132& 108& 227\\
                                                                                      &&&&&&&&&&&                         &4340& -634& 660& 240& 411& 256& 326& 276& 235& 290& 184& 256& 135& 222& 90& 152& 67& 17& 318\\
                                                                                                                &&&&&&&&&&&&      &2986& -514& 479& 340& 363& 377& 362& 315& 343& 265& 311& 144& 198& 17& 62& 86& 147& 226\\
                                                                                  &&&&&&&&&&&&&                                              &3592& -134& 606& 333& 422& 374& 333& 349& 267& 300& 157& 184& 9& 71& 85& 136& 202\\
                                                                                          &&&&&&&&&&&&&&                                                   &1401& 22& 431& 343& 349& 302& 322& 245& 284& 171& 186& 70& 70& 93& 142& 202\\
                                                                                                                           &&&&&&&&&&&&&&&                          &1491& 141& 506& 386& 356& 394& 278& 306& 188& 212& 79& 71& 106& 145& 240\\
                                                                              &&&&&&&&&&&&&&&&                                                                                &1203& 200& 435& 331& 379& 281& 311& 184& 209& 49& 51& 110& 197& 181\\
                                                                                                                   &&&&&&&&&&&&&&&&&                                               &1032& 258& 408& 398& 305& 330& 197& 223& 78& 79& 113& 174& 225\\
                                                                                                                                                           &&&&&&&&&&&&&&&&&&                  &1086& 232& 453& 298& 328& 120& 189& -48& 22& 42& 142& 204\\
                                                                                            &&&&&&&&&&&&&&&&&&&                                                                                            &1006& 151& 329& 282& 169& 195& 58& 80& 95& 192& 188\\
                                                                                         &&&&&&&&&&&&&&&&&&&&                                                                                                           &1541& 124& 400& 199& 261& 150& 166& 202& 251& 251\\
                                                                                                  &&&&&&&&&&&&&&&&&&&&                                                                                                        &  &1127& 72& 227& 222& 93& 118& 93& 171& 161\\
                                                                                                                                                    &&&&&&&&&&&&&&&&&&&& &                                                      &&1723& -105& 406& -3& 180& 190& 198& 247\\
                                                                                                                                                                                                       &&&&&&&&&&&&&&&&&&&& &&                &      &1550& 144& 946& 502& 647& 437& 215\\
                                                                                                                                                                 &&&&&&&&&&&&&&&&&&&& &&&                                                      &             &1292& 187& 524& 393& 387& 284\\
                                                                                                                                                                  &&&&&&&&&&&&&&&&&&&& &&&&                                                            &                    &3941& 587& 1657& 641& 346\\
                                                                                                                                                                     &&&&&&&&&&&&&&&&&&&& &&&&&                                                                           &           &2980& 360& 1124& 305\\
                                                                                                                                                                                                                                 &&&&&&&&&&&&&&&&&&&& &&&&&&                        &           &4465& -1891& 713\\
                                                                                                                                                                               &&&&&&&&&&&&&&&&&&&& &&&&&&                                                           &      &        &23902& -1826\\
                                                                                                                                                                                                                                              &&&&&&&&&&&&&&&&&&&& &&&&&&&&&                            &19169\end{pmatrix},
$}
\end{gather*}

Also, $\mu_b$ stands for the observational distance modulus, $M$ is a
free normalization parameter which should be constrained and the dimensionless
luminosity distance is defined as

\begin{equation}\label{dL}
d_L=\frac{c\left(1+z\right)}{H_0}\int_{0}^{z'}
\frac{dz'}{H\left(z\right)}.
\end{equation}
 For Baryon Acoustic Oscillations (BAO), we use the BOSS DR12 including six data points \cite{78}. The $\chi^2_{BAO}$ function can be explained as
\begin{equation}\label{BAO}
\chi^2_{BAO}=X^tC_{BAO}^{-1}X,
\end{equation}
where $X$ for six data points will be
\begin{equation}\label{XBAO}
X=\left(\begin{array}{c} \frac{D_M\left(0.38\right)r_{s,fid}}{r_s\left(z_d\right)}-1512.39\\
\frac{H\left(0.38\right)r_s\left(z_d\right)}{r_s\left(z_d\right)}-81.208\\
\frac{D_M\left(0.51\right)r_{s,fid}}{r_s\left(z_d\right)}-1975.22\\
\frac{H\left(0.51\right)r_s\left(z_d\right)}{r_s\left(z_d\right)}-90.9\\
\frac{D_M\left(0.61\right)r_{s,fid}}{r_s\left(z_d\right)}-2306.68\\
\frac{H\left(0.51\right)r_s\left(z_d\right)}{r_s\left(z_d\right)}-98.964\end{array}\right),
\end{equation}
and $r_{s,fid}=$147.78 Mpc is the sound horizon of fiducial model, $D_M(z)=(1+z)D_A(z)$ is the comoving angular diameter distance. The sound horizon at the decoupling time $r_s(z_d)$ is defined as

\begin{equation}\label{BAO}
r_s\left(z_d\right)=\int_{z_d}^{\infty} \frac{c_s\left(z\right)}{H\left(z\right)}dz,
\end{equation}
in which $c_s=1/\sqrt{3(1+R_b/(1+z))}$ is the sound speed with
$R_b=31500\Omega_bh^2(2.726/2.7)^{-4}$. The covariance matrix $Cov_{BAO}$ \cite{78} is:
\begin{equation}\label{iCOVBAO}
C^{-1}_{BAO}=10^{-4}\begin{pmatrix}   624.707& 23.729  &325.332    &8.34963&   157.386 &3.57778\\
23.729  &5.60873    &11.6429    &2.33996    &6.39263    &0.968056\\
325.332 &11.6429    &905.777    &29.3392    &515.271&   14.1013\\
8.34963 &2.33996    &29.3392    &5.42327    &16.1422&   2.85334\\
157.386 &6.39263    &515.271    &16.1422    &1375.12&   40.4327\\
3.57778 &0.968056   &14.1013    &2.85334    &40.4327
&6.25936\end{pmatrix}.
\end{equation}
Studying the expansion time line of the universe, we check Cosmic Microwave Background(CMB). For this, we use Planck 2015 data set \cite{79}. The $\chi^2_{CMB}$ function may be explained as
\begin{equation}\label{CMB}
\chi^2_{CMB}=q_i-q^{data}_i Cov^{-1}_{CMB}\left(q_i,q_j\right),
\end{equation}
where $q_1=R(z_*)$, $q_2=l_A(z_*)$ and $q_3=\omega_b$ and $Cov_{CMB}$
is the covariance matrix \cite{79}. The datapoints of Planck 2015 are
\begin{equation}\label{PLANCKDATA}
q^{data}_1=1.7382,~\\
q^{data}_2=301.63,~\\
q^{data}_3=0.02262.
\end{equation}
The acoustic scale $l_A$ is
\begin{equation}\label{lA}
l_A=\frac{3.14d_L\left(z_*\right)}{\left(1+z\right)r_s\left(z_*\right)},
\end{equation}
in which $r_s(z_*)$ is the comoving sound horizon at the drag epoch ($z_*$). The function of redshift at the drag epoch is \cite{80}

\begin{equation}\label{z_*}
z_*=1048\left[1+0.00124\left(\Omega_bh^2\right)^{-0.738}\right]\left[1+g_1\left(\Omega_mh^2\right)^{g_2}\right],
\end{equation}
where
\begin{equation}\label{g1 g2}
g_1=\frac{0.0783\left(\Omega_bh^2\right)^{-0.238}}{1+39.5\left(\Omega_bh^2\right)^{-0.763}}, ~~~g_2=\frac{0.560}{1+21.1\left(\Omega_bh^2\right)^{1.81}}.
\end{equation}
The CMB shift parameter is \cite{81}
\begin{equation}\label{R}
R=\sqrt{\Omega_{m_0}}\frac{H_0}{c}r_s\left(z_*\right).
\end{equation}
For studying the cosmic expansion history, the specification of Hubbe parameter using observational data is of utmost importance . The $\chi^2_{OHD}$ is
\begin{equation}\label{OHD}
\chi^2_{OHD}=\sum_{n=1}^{i}\frac{[H_{obs}\left(z_i\right)-H_{th}\left(z_i\right)]^2}{\sigma^2_i}
\end{equation}
We use 43 data points  in the redshift range $0<z<2.5$ \cite{82}

Eventually, the $\chi^2$ for SN Ia, BAO and CMB is
\begin{equation}\label{TOTALX}
\chi^2_{total}=\chi^2_{SNIa}+\chi^2_{BAO}+\chi^2_{CMB}+\chi^2_{OHD}.
\end{equation}
Using minimized $\chi^2_{total}$, we can constrain and obtain the best fit values of the free parameters. The best-fit values of $\Omega_m$, $ H_0$, $ \beta$, c, $b^2 $, $ r_c$ and M by consideration of the $1\sigma$ and $2\sigma$ confidence level are shown in the Table 1.

\begin{table}[h]
\begin{center}
\begin{tabular}{ccc}
\hline\hline Parameters & cJLA + BOSS DR12 + Planck2015+ OHD  \\ \hline\hline
 \multicolumn{1}{c|}{$\Omega_m$ }                           & \multicolumn{1}{c}  {$0.278^{+0.007~+0.012}_{-0.008~-0.011}$}\\  [0.1cm]\hline
 \multicolumn{1}{c|}{$H_0$ }                           & \multicolumn{1}{c}  { $69.9^{+0.95~+1.57}_{-0.95~-1.57}$}\\ [0.1cm] \hline
 \multicolumn{1}{c|}{$\beta$ }                           & \multicolumn{1}{c}  {$0.496^{+0.005~+0.009}_{-0.005~-0.009}$}\\ [0.1cm] \hline
 \multicolumn{1}{c|}{$r_c$ }                           & \multicolumn{1}{c}  {$0.08^{+0.02~+0.027}_{-0.02~-0.027}$}\\  [0.1cm]\hline
 \multicolumn{1}{c|}{$c$ }                           & \multicolumn{1}{c}  {$0.691^{+0.024~+0.039}_{-0.025~-0.037}$}\\  [0.1cm]\hline
 \multicolumn{1}{c|}{$b^2$ }                           & \multicolumn{1}{c}  { $0.035$}          \\ [0.1cm] \hline
 \multicolumn{1}{c|}{$M$ }                           & \multicolumn{1}{c}  {$10.7$}\\ [0.1cm]
\hline\hline
\end{tabular}
\caption{68.3\% and 95.4\% error marginalized result for each
parameter.}
\end{center}
\end{table}
 \begin{figure}[H]
\begin{center}
\includegraphics[width=0.8\textwidth]{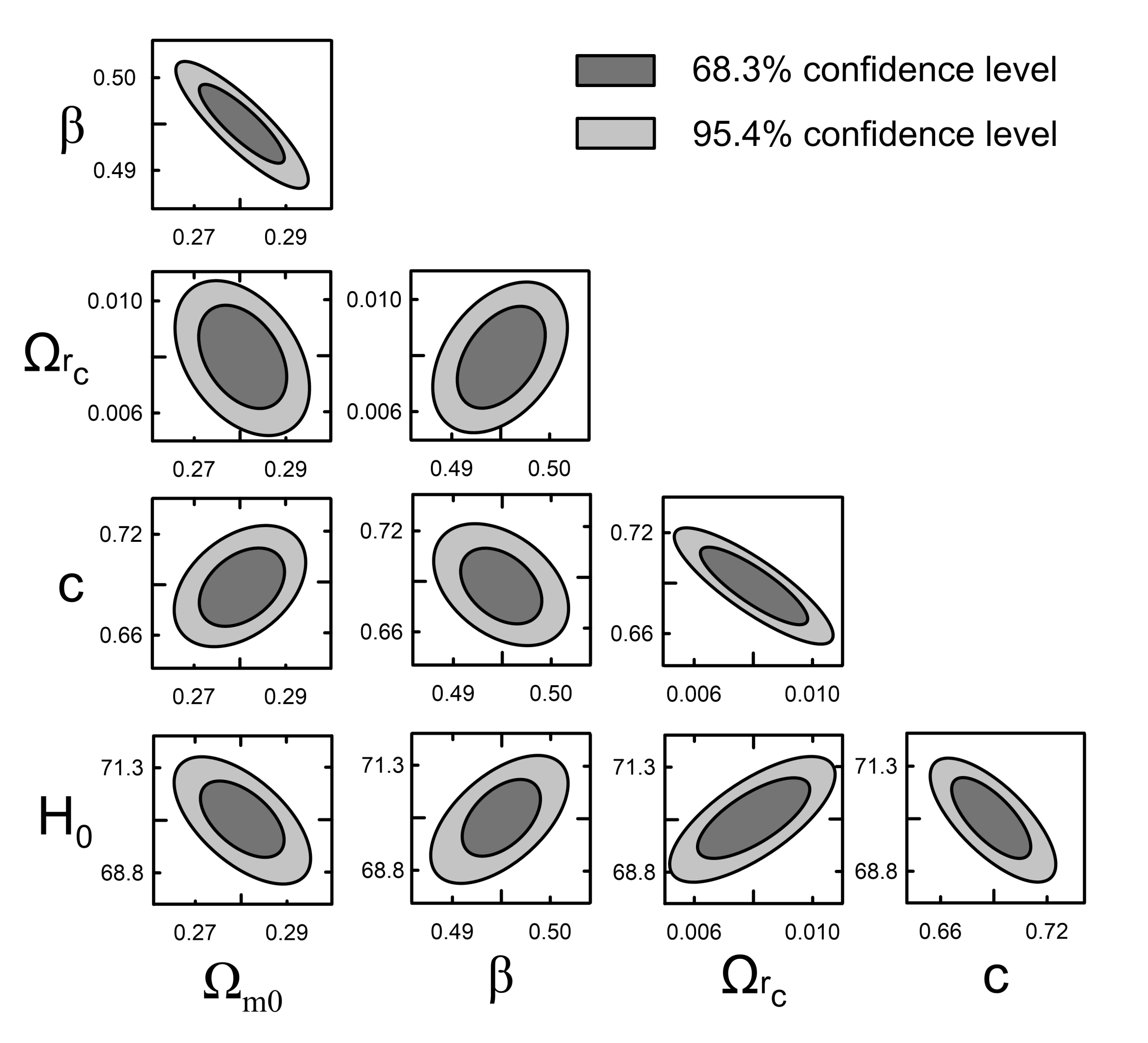}
\caption{The 2D 68.3$\%$ and 95.4$\%$ confidence level for $H_0$, $\Omega_{m0}$, c, $\Omega_{r_c}$ and $\beta$.} \label{fig7}
\end{center}
\end{figure}
Using the latest observational data sets, we have plotted 2D confidence region of the important parameters
of the current model in Fig. \ref{fig7}.

    \section{CONCOLUSIONS}
  In this paper, we studied a new holographic dark energy model
with Hubble horizon as IR cutoff in fractal universe for flat FRW space. The features of fractal cosmology could remove ultraviolet divergencies and also make a better understanding of the universe in different dimensions. There are many models of holographic dark energy created by cosmologists with various cosmological constraints and explain that the size of $L$ as the horizon length should be less than the mass of black hole with the same size. In order to prevent formation of black hole, the minus sign of the equation of state of dark energy is not enough. Phantom-like dark energy is the most appropriate condition to avoid the formation of black hole ($\omega_D<-1$). In the present work, calculation of the EoS parameter for NHDE with Hubble horizon as IR cutoff demonstrated that the EoS parameter lies in phantom realm for interacting mode ($b^2=0.035$-fitted with recent observational data) as seen in Fig. \ref{fig2}. Plotting the deceleration parameter showed an accelerating expansion for the universe and behaves approximately similar to the cold dark matter.
The plot of the $Om$-diagnostic tool against redshift by taking $x = ln(1+z)^{-1}$ is shown in Fig. \ref{fig4}.
It may be noted that the choice of HDE in this work is a very
particular example of generalized HDE introduced in Nojiri and
Odintsov, where a phantom cosmology based approach towards
unification of early and late-time universe was proposed.

The positive value of the trajectories in the $Om$-diagnostic plane in the late time for all values of the coupling constant can be observed, which indicates the phantom behavior and shows a suitable uniformity with the equation of state parameter obtained by observational data. The $r-s$ StateFinder plane for the present framework is plotted in Fig. \ref{fig5}. We can see that all trajectories for all cases of $b^2$ in the $r-s$ plane meet the $\Lambda$CDM fixed point $(r,s)=(1,0)$. The StateFinder trajectories indicate the Chaplygin gas behavior for NHDE (where $s<0$, $r>1$) and also it shown a phantom-like behavior ($s<0$). This is consistent with the equation of state results. For further information, the main model of NHDE\cite{8} in the simple FRW universe could hardly reach the phantom area and according to the Eos parameter, its behavior is similar to quintessence DE. We, also have studied the future singularities for NHDE in the
framework of fractal cosmology for two types of interaction term
$Q$ considering the Hubble parameter as
$H=(\frac{h}{t_0-t})^\beta+H_0$. For $Q$ to be
proportional to the energy density, it is observed that the model
is characterized by Type I i.e. Big Rip singularity. However, for
$Q$ proportional to matter density, the model is found to have
asymptotic de Sitter solution corresponding to the so called
w-singularity. In this connection it may be noted that an explicit
cosmological model involving w-singularity was proposed in
having finite scale factor, vanishing energy density and pressure,
and the only singular behavior appears in the barotropic index
$w(t)$.

Finally, in order to check compatibility with observational data
and fitting the free parameters, we used cJLA compilation for
SNIa, six observational points of BAO from BOSS DR12, Planck 2015
for CMB and current observational data points for OHD. Using MCMC method and the combination of the latest observational data sets we obtained
 $\Omega_{m0}=0.278^{+0.008~+0.010}_{-0.007~-0.009}$,
$H_0=69.9^{+0.95~+1.57}_{-0.95~-1.57}$,
$r_c=0.08^{+0.02~+0.027}_{-0.02~-0.027}$,
$\beta=0.496^{+0.005~+0.009}_{-0.005~-0.009}$,
 $c=0.691^{+0.024~+0.039}_{-0.025~-0.037}$ and $b^2=0.035$ with
$1\sigma$ and $2\sigma$ confidence interval. While concluding, the
similar analysis is proposed for other generalizations of HDE as
future work.

      \section*{\centering \small ACKNOWLEDGMENTS }
  We would like to thank the referee for insightful comments which improved the quality of the paper. Martiros Khurshudyan is supported in part by Chinese Academy of Sciences President’s International Fellowship Initiative Grant (No. 2018PM0054). Surajit Chattopadhyay is financially supported by CSIR Grant 03(1420)/18/EMR-II.

 \section*{\centering \small COMPLIANCE WITH ETHICAL STANDARDS}
The authors hereby ensure that the accepted principles of ethical and professional conduct have been followed.

    \bibliographystyle{unsrt}

  \end{document}